\documentclass[12pt]{article}
\usepackage{amsmath}
\usepackage{amssymb}
\usepackage{pstricks}
\usepackage{stmaryrd}
\usepackage{graphicx}
\usepackage{pst-node}

\newcommand{\be}{\begin{equation}}
\newcommand{\ee}{\end{equation}}
\newcommand{\ba}{\begin{eqnarray}}
\newcommand{\ea}{\end{eqnarray}}

\newcommand{\bd}{\begin{description}}
\newcommand{\ed}{\end{description}}

\newtheorem{lemma}{Lemma}
\newtheorem{theorem}{Theorem}
\newtheorem{proposition}{Proposition}
\newtheorem{corollary}{Corollary}

\newcommand{\nn}{\nonumber}

\renewcommand{\iota}{{\bf 1}}
\newcommand{\opsi}{\overline{\psi}}
\newcommand{\e}{\epsilon}
\newcommand{\z}{\mathfrak{z}}
\renewcommand{\l}{l}

\def\rellow#1#2{Mathrel{Mathop{\kern 0pt #1}\limits_{#2}}}

\def\H{{\cal H}}

\def\U{{\cal U}}

  \newcommand{\VR}{V_{R}}
  \newcommand{\VRtilde}{\tilde{V}_{R}}
  \newcommand{\VL}{V_{L}}
  \newcommand{\Cdrei}{\mathbb{C}^{3}}
  \newcommand{\MD}{M_{D}}
  \newcommand{\MU}{M_{U}}

  \newcommand{\upfrd}{\ar@<0.3ex>@{.>}[d]}
  \newcommand{\pfr}{\ar@<0.3ex>[r]}
  \newcommand{\pfl}{\ar@<0.3ex>[l]}
  \newcommand{\upflu}{\ar@<0.3ex>@{.>}[u]}

\title{Gauge Theories\\ with Nonunitary Parallel Transporters: \\
A soluble Higgs model
\footnote{Work supported by Deutsche Forschungsgemeinschaft}
  }

\author{Claudia Lehmann and Gerhard Mack\\
   II. Institut f\"ur Theoretische Physik, Universit\"at Hamburg
\date{\today}}
\begin{document}
\maketitle
$\quad$\\[-10mm]\noindent 
{\bf abstract}
It has been proposed to abandon the requirement that parallel transporters in gauge theories are unitary (or pseudoorthogonal). 
This leads to a geometric interpretation of Vierbein fields as parts of gauge fields, and nonunitary parallel transport in extra directions yields Higgs fields. In such theories, the holonomy group $H$ is larger than the gauge group $G$.  Here we study a 1-dimensional model with fermions which retains only the extra dimension, and which is soluble in the sense that its renormalization group flow may be exactly computed, with $G=SU(2)$ and noncompact $H\subseteq GL(2,\mathbf{C})$, or $G=U(2), H=GL(2,\mathbf{C})$. In all cases the asymptotic behavior of the Higgs potential is computed, and with one possible exception for $G=SU(2)$, $H=GL(2,\mathbf{C})$, there is a flow of the action from 
 an UV-fix point which describes a $SU(2)$-gauge theory with unitary parallel transporters, to a IR-fixpoint.
 We explain how a splitting of the masses of fermions of different flavor can arise through spontaneous symmetry breaking. 

\newpage
\section{Introduction}\label{intro}
In gauge theories, parallel transporters $\U(C)$ along
 paths $C$ are required by the principles, and are 
determined by vector potentials. 
The lattice gauge fields of lattice gauge theory are parallel transporters. Traditionally they are unitary maps
between vector spaces with scalar products.  
It has been proposed \cite{generalGaugeI} to abandon the additional requirement of their unitarity (or pseudoorthogonality), $\U(C)^\ast=\U(C)^{-1}$. 
This leads to a geometric interpretation of Vierbein fields as parts of gauge fields, and nonunitary parallel transport in extra directions yields Higgs fields. One gets, besides the unitary gauge group $G$, a larger holonomy group $H\supset G$ with an involutive automorphism $\theta$ such that $\theta(g)=g$ if and only if $g\in G$. Only $G$ is a local symmetry, but fields form representation spaces of $H$. In lattice models, the lattice gauge fields take their values in $H$. Here we study a 1-dimensional model with fermions which retains only the extra dimension, and which is soluble in the sense that its renormalization group flow may be exactly computed, with $G=SU(2)$, and $H\subseteq GL(2,\mathbf{C})$.

We consider a 1-dimensional Euclidean lattice model with fermions. It lives on a chain of lattice spacing $a_0=1$ whose sites are labeled by $z\in \mathbf{Z}$. 

The Fermi fields $\psi (z), \bar{\psi}(z)$ sit on the sites of the chain $\Lambda^0$ . They have only one spin component, but two color-components. Elements $\phi$ of $H\subseteq GL(2,\mathbf{C})$ can act on them. They transform according to the fundamental representation  of the gauge group $G=SU(2)$. As usual, $\psi$ and $\bar{\psi}$ are treated as independent integration variables in the Feynman Kac path integral \cite{MuensterMontvay} .  

With every link $(z,z+1)$ and its adjoint $(z+1,z)$ between neighboring sites there is associated a possibly nonunitary lattice gauge field (parallel transporter) $\phi(z+1,z)\in H$ and its adjoint,
\be \phi(z,z+1)=\phi(z+1,z)^\ast. \nn \label{adjPhi}\ee

The groups of interest are $G=SU(2)$, 
\be H=\mathbf{R}_+SU(2), \ SL(2,\mathbf{C}), \mathbf{R}_+SL(2,\mathbf{C}) ,\ GL(2,\mathbf{C}) \nn.
\ee
In section \ref{sec:u2} the results will be extended to 
\be G=U(2)\ , H=GL(2,\mathbf{C}). \nn \ee

Under a $SU(2)$ gauge transformation,
\ba \psi(z)&\mapsto& u(z)\psi(z) \nn \\
\phi(z+1) &\mapsto& u(z+1)\phi(z+1,z)u(z)^{-1}. \nn
\ea 
The initial action is 
\ba S^0(\psi , \bar{\psi}, \phi)&=&\sum_z 
\Big(S^0_F(z,z+1)  -  V(\phi(z,z+1)) \Big), 
 \nn \\
S^0_F(z,z+1) &=& 
\z \bar{\psi}(z)\phi(z,z+1)\psi(z+1) + 
\bar{\z}\bar{\psi}(z+1)\phi(z+1,z)\psi(z) .
 \label{originalAction} 
\ea
The constant factor $\z$ and its complex conjugate $\bar{\z}$ are included for later convenience.
 $V(\Phi )$ is interpreted as a Higgs potential. It is a real function of $\Phi \in H$ with the following further property, which will be shared by not necessarily real coefficients in the effective actions,
\be 
V(\Phi^\ast) = \overline{V(\Phi)} \label{Vreality}.
\ee
The action is required to be invariant under $SU(2)$-gauge transformations. Therefore 
$$V(u_1\Phi u_2)=V(\Phi)\ \mbox{ for all } u_1,u_2\in SU(2),\ \Phi\in H.$$
 The path integral involves the Haar measure 
$d\Phi$ on $H$ to integrate over the lattice gauge fields $\phi(z+1,z)$.
We assume that $e^{-V(\Phi)}$ is integrable and square integrable on the group $H$, hence
$$ V(\Phi)\mapsto +\infty \mbox{ when } ||\Phi ||\mapsto \infty . $$

 The initial Boltzmann factor is
\be
B^0(\psi , \bar{\psi}, \phi) = e^{S^0} = \prod_z B^0(z,z+1) = \prod_zB_F^0(z,z+1)e^{-V^0(\phi(z,z+1))} \nn
\ee
where $B^0(z,z+1)$ depends on $\psi(z),\opsi(z),\psi(z+1),\opsi(z+1),\phi(z,z+1)$. 

For the groups studies here, $\Phi\in H$ can be parametrized as $\Phi = u_1d(\eta)u_2$, $u_i\in G$, and $d(\eta)$ a diagonal matrix depending on a parameter $\eta$ in a linear space $\mathfrak{h}$. Hence
\be
V(\Phi)= \mathcal{V}(\eta) \ 
\ee
by $G$-biinvariance. Typically there is a discrete subgroup $W_T\subset G$ with elements $w$  such that 
$w d(\eta)w^{-1} = d(w \eta)$ for some nontrivial action of $w$ on $\mathfrak{h}$. It follows that $\mathcal{V}(\eta)= \mathcal{V}(w \eta)$. We will show in section \ref{sec:masssplit} that flavor mass splitting results when this symmetry under $W_T$ is broken spontaneously, i.e. if the orbits of minima of $\mathcal{V}$ do not consist of a single point.  

In the rest of this paper we study the real space renormalization group (RG) flow of this 1-dimensional model. After $n$ renormalization group steps, the model lives on a chain $\Lambda^n \subset \Lambda^0$ of lattice spacing $a_n=L^n a_0$, and depends on fields $\psi^n , \bar{\psi}^n$ on the sites of the block lattice $\Lambda^n$ and on lattice gauge fields $\phi^n(z,z+a_n)$ sitting on the links. Somewhat surprisingly, it turns out that the block side $L$ has to be chosen odd, and we choose $L=3$. For $L=2$, the effective Boltzmann factor would not admit taking the logarithm to obtain the effective action, see section \ref{sec:3.2}.

The block spin transformation will act on the fermions by decimation, so that
\be \psi^n(z)=\psi(z) \mbox{ for } z\in \Lambda^n \label{decim} \nn\ee
and similarly for $\opsi$. 
For the gauge fields, the appropriate choice of block spin is 
\be
\phi^{n+1}(z+3a_n,z)= \phi^n(z+3a_n,z+2a_n)\epsilon
 \phi^n(z+a_n,z+2a_n)^T\epsilon^T \phi^n(z+a_n,z) , \label{Phi}
\ee
for $n=0,1,2,...$, with $\phi^0=\phi$, $a_0=1$. The expression involves the antisymmetric tensor $\epsilon$ in two dimensions, with $\epsilon_{12}=1$, and $A^T $ stands for the transpose of the matrix $A$.  In particular
\be
\phi^1(z+3,z)= \phi(z+3,z+2)\epsilon
 \phi(z+1,z+2)^T\epsilon^T \phi(z+1,z) , \label{Phi1}
\ee
Under the RG-transformation steps considered here, the effective actions will remain strictly local in the sense that they are sums of contributions from block links,
\be S^n(\psi^n, \opsi^n,\phi^n) = \sum_{z\in \Lambda^n} S^n(z,z+a_n) 
\label{effSum} \nn \ee
with $S^n(z,z+a_n)$ depending only on $\psi^n(z), \opsi^n(z),\psi^n(z+a_n), \opsi^n(z+a_n),$ and $\phi^n(z, z+ a_n)$. 

When $H$ is not a subgroup of $SL(2,\mathbf{C})$, the effective actions are more complicated than (\ref{originalAction}). Integration of the Higgs fields $\phi$ produces 4-fermion, 6-fermion and 8-fermion interactions,
with $\phi$-dependent coefficients. We present recursion relations for the corresponding coefficients, and their solution. The coefficients in the effective action are obtained by taking the logarithm. 
In case $H=SL(2,\mathbf{C})$ there is a great simplification, so that this model is nearly trivial. The effective action retains the form 
(\ref{originalAction}), and the flow of the effective Higgs potential $V$ receives no contributions form the fermions. 
\section{Generation of flavor mass splitting}
\label{sec:masssplit}
For definiteness consider $H=SL(2,\mathbf{C})$, $G=SU(2)$. Matrices $A\in  SL(2,\mathbf{C})$ may be parametrized as $A=u_1d(\eta)u_2$ with $u_i\in G$, $\eta$ real and
$d(\eta)=diag(e^{-\eta/2}, e^{\eta/2})$. There is $w\in G$ such that
 $w d(\eta)w^{-1}=d(-\eta)$, 
cp. Lemma \ref{lemma:2} of section \ref{sec:gaugeinv}. Therefore $\mathcal{V}(\eta)=V(d(\eta))$ obeys 
$$\mathcal{V}(\eta) = \mathcal{V}(-\eta) . $$  
Suppose that the orbit of the minima under the discrete group $W_T=\{ 1 , w \}$ is nontrivial, then the minima of $\mathcal{V}(\eta)$ are at $\pm \hat{\eta}\neq 0$. Then at the minima of $V(\phi)$, $\phi = u_1d(\hat{\eta})u_2$ and the interaction term becomes
\be
\z \overline{\Psi}\phi \Psi = \sum_{\alpha=1}^2 
\overline{\Psi}_\alpha^\prime m_\alpha \Psi_\alpha^\prime \ 
\ee
with $\overline{\Psi}^\prime = \overline{\Psi}u_1,$
$\Psi^\prime=u_2\Psi$, and masses
$$ m_1=\z e^{-\hat{\eta}/2}\ ,\quad m_2=\z e^{\hat{\eta}/2} \ .$$
We see that we have fermions of two flavors with different masses.

Although the 1-dimensional model does not show this feature, we hope that in realistic models there is a RG-flow from infinitely deep single well $\mathcal{V}$ as corresponds to ordinary gauge theory to $\mathcal{V}$ with multiple minima. If the RG-flow is stopped there because the extra dimension has finite extent, we get flavor mass splitting.  
\section{Summary of results on the RG-flow}\label{section:Summary}
\subsection{General form of the effective action}
 In case $H\neq SL(2,\mathbf{C})$, non-bilinear fermion actions will be produced by the RG-flow. 
It follows from Fierz identities, Lemma \ref{lemma:7}, that the most general action has the form
\ba
S^n(z,z+a_n) &=& c_0^n + c_1^n L^n  + \bar{c}_1^n \bar{L}^n\nn \\
&&
  + c^n_2 k^n + \bar{c}^n_2 \bar{k}^n + c^n_3 L^n\bar{L}^n \nn \\
&&
 + c^n_4 k^n\bar{L^n} + \bar{c}^n_4 \bar{k}^nL^n 
+ c^n_5k^n\bar{k}^n \ . \label{generalAction} 
\ea
where 
\ba
L^n &=& \opsi(z)\phi^n(z,z+a_n)\psi(z+a_n) \ , \label{def_L}\\
\bar{L}^n &=&  \opsi(z+a_n)\phi^n(z+a_n,z)\psi(z) \ , \nn\\
k^n &=& \frac 1 4 \opsi\e\opsi(z)\times\psi\e\psi(z+a_n)
\ , \label{def_k}\\
\bar{k}^n &=& \frac 1 4 \opsi\e\opsi(z+a_n)\times\psi\e\psi(z)\ .  \nn
\ea
and where $c_i^n$ are functions of $\phi^n(z,z+a_n)=\phi^n(z+a_n,z)^\ast$; $\bar{c}_i^n$ is the complex conjugate of $c_i^n$, and $c_0^n$, $c_3^n$ and $c_5^n$ are real. In particular, $c_0^n$ is the negative effective Higgs potential
\be
c_0^n = -V^n\big(\phi^n(z,z+a_n)\big)\nn .  
\ee
We identify $V^0=V$.

The Boltzmann factor will have the same general form,
\ba
B^n(z,z+a_n) &=& a_0^n + a_1^n L^n  + \bar{a}_1^n \bar{L}^n \nn \\
&&   + a^n_2 k^n + \bar{a}^n_2 \bar{k}^n + a^n_3 L^n\bar{L}^n \nn \\
&&  + a^n_4 k^n\bar{L}^n + \bar{a}^n_4 \bar{k}^nL^n
+ a_5^nk^n\bar{k}^n \label{generalBoltzmann}
\ea
with $a_0^n=\exp \Big(-V^n\big(\phi^n(z,z+a_n)\big)\Big)$. 

In more careful notation we write
\ba
a_i^n &=&a_i^n(z,z+a_n) = a_i^n[\phi^n(z,z+a_n)] \ , \nn \\ 
\bar{a}_i^n &=&\bar{a}_i^n(z,z+a_n) = \overline{a_i^n[\phi^n(z,z+a_n)]} \nn .
\ea
The following reality requirements will be satisfied,
\be a^n_j[\Phi^\ast] = \bar{a}_j^n[\Phi] \label{aReality} \nn
\ee
and similarly for the coefficients $c^n_j$ of the action. It follows from eq.(\ref{Vreality}) that they are fulfilled for the initial values $a^0_j$, and they will pass through the recursion relations. 
%
\subsection{Flow of the effective action for general $H$}
\begin{theorem} \emph{(RG-flow for general $H$)} \label{theorem:3} \\
The coefficients $a_i^n[\Phi]$ in the effective Boltzmann-factor obey the following recursion relations,
\ba
a_0^{n+1}[\Phi ]&=& a_0^n*\bar{a}_5^n*a_0^n [\Phi]\ , \nn \\ 
a_1^{n+1}[\Phi ]&=& a_1^n *\bar{a}_4^n\ast a_1^n [\Phi ]
 \ ,\nn \\
a_2^{n+1}[\Phi ]&=& a_2^n\ast \bar{a}_2^n\ast a_2^n [\Phi ] \ , \nn \\
 a_3^{n+1}[\Phi ]&=& a_3^n \ast \bar{a}_3^n \ast a_3^n [\Phi ] \ , \nn \\
 a_4^{n+1}[\Phi ]&=& a_4^n \ast \bar{a}_1^n\ast a_4^n [\Phi ] , \ 
 \nn \\
 a_5^{n+1}[\Phi ]&=& a_5^n \ast \bar{a}_0^n \ast a_5^n [\Phi ] \  ,\nn 
\ea
where $\ast$ is the convolution product on the group $H$,
and $\Phi\in H$.  $(\bar{a}_i^n ={a}_i^n$ for 
$i=0,3,5).$ 

All the coefficients are $SU(2)$-biinvariant functions on $H$, viz.
$ a_i[u_1\Phi u_2] = a_i[\Phi] $
for $\Phi \in H$, $u_1, u_2\in SU(2).$

The coefficients $c^n_i$ in the effective action are related to $a^n_i$ by proposition \ref{proposition:takeLog} of section \ref{sec:takeLog}.
\end{theorem}
 We may turn the convolution product into a product by harmonic analysis on the group $H$. 

For our groups $H$, every unitary irreducible representation contains at most one normalized $SU(2)$-invariant vector in its representation space.  Let us denote by $\Xi$ those representations which possess such a vector, and which contribute to the harmonic expansion of good functions (=square integrable on the group), and let us denote the restriction of the Plancherel measure 
to such representations by $d\Xi$.

Let us denote the corresponding matrix element of the representation operator by $P^\Xi(\Phi)$, and call it spherical function.
Good $SU(2)$-biinvariant functions on $H$ admit an expansion in spherical functions, $
f[\Phi] = \int d\Xi \tilde{f}(\Xi )P^\Xi(\Phi ) \ .
$ 
Under harmonic analysis, convolutions turn into products. Therefore we have the following corollary of the recursion relations
\begin{corollary} \label{corollary:1}
Write $\bar{a}_i = a_{\bar{i}}.$ Then,
\be
a_i^{n+1}[\Phi ]= \int d\Xi \ \tilde{a}_i^{n+1}(\Xi) \ P^\Xi(\Phi)\ , \nn \ee
with Fourier coefficients
\be
\tilde{a}_i^{n+1}(\Xi) = \sum_{jkl}d^i_{\ jkl}\tilde{a}_j^n(\Xi)\tilde{a}_k^n(\Xi)\tilde{a}_l^n(\Xi) \ ,\nn
\ee
where
\be
\tilde{a}_i^n(\Xi) = \int_H d\Phi \ a_i^n [\Phi]P^\Xi(\Phi^{-1}) \label{inv} 
\nn 
\ee
The sum over $j,k,l= 0,1,\bar{1},2,\bar{2},3,4,\bar{4},5$ has only one nonvanishing term, which is equal to $1$. The nonvanishing coefficients are 
\be  d^0_{\ 050}, \ d^1_{\ 1\bar{4}1},\ d^2_{\ 2\bar{2}2},\ d^3_{\ 333},
\ d^4_{\ 4\bar{1}4}, \ d^5_{\ 505}; \quad
  d^{\bar{1}}_{\ \bar{1}4\bar{1}}, \  d^{\bar{2}}_{\ \bar{2}2\bar{2}}, \   d^{\bar{4}}_{\ \bar{4}1\bar{4}}.\nn
\ee 
\end{corollary} 
The explicit description of $\Xi$ and $P^\Xi$ for the various groups $H$ is found in Appendix A.

The result is also valid for $H=SL(2,\mathbf{C})$. It permits to compute the renormalization group flow when one starts with a more general action than (\ref{originalAction}).

\subsection{Flow between fix points}

The asymptotic behavior of the Higgs potential after many
$RG$-steps can be computed. It becomes flat in the case
$\mathbf{R}_+SU(2)$, and tends to a definite function on
$H$ in the other cases. This is also true 
for the model with $G=U(2), H=GL(2,\mathbf{C})$.

There is a UV-fix point which describes a $SU(2)$-gauge theory with fermions with unitary parallel transporters $\phi(z,z+1)\in SU(2)$. 
\be
e^{-V(\Phi)} = \delta_{SU(2)}(\Phi) \ . \label{UV-fixpoint}
\ee
 $\delta_{SU(2)}(\Phi)$ is a $\delta$-function concentrated on $SU(2)\subset H$, so that \\
 $\int_H d\Phi f(\Phi)\delta_{SU(2)}(\Phi)=
 \int_{SU(2)} du f(u).$
Since the unitary parallel transporters may be transformed away in 1 dimension, this model is equivalent to a free fermion theory. 
The asymptotic behavior of the model after a large number $n$ of RG-steps may be computed, starting from 
$e^{-V(\zeta d(\eta))}$  which may be arbitrarily close to the UV-fix point (\ref{UV-fixpoint}) and satisfies the following assumptions. It is square integrable on $H$, piecewise continuous, and $\int d\Phi |\zeta|^\epsilon e^{\epsilon |\eta|}e^{-V} < \infty$ for some $\epsilon > 0$. 

In all cases, including $G=U(2), H=GL(2,\mathbf{C})$ but with the possible exception of $G=SU(2), H=GL(2,\mathbf{C})$,
there is a flow away from this fix point and towards a 
IR-fix point. In the exceptional case we are unable to 
compute the asymptotic behavior of some coefficients in the fermionic action other than the Higgs potential. 

  The case $G=SU(2), H=GL(2,\mathbf{C})$ is different from all the others in that $G$ is not the maximal compact subgroup of $H$.

\subsubsection*{$\mathbf{H=SL(2,C)}$}
We use the parametrization $A=u_1d(\eta)u_2$ of elements of $A\in SL(2,\mathbf{C})$, that will be given explicit in section \ref{sec:3.4}.
 
Choose the additive constant in $V$ such that 
$\int_{SL(2,\mathbf{C})} e^{-V(A)}dA = 1$. Let
\be
b=\int  \frac{1}{12} \sinh^2\eta d\eta \eta^2 e^{-V(d(\eta))}\ \nn
\ee
and $B=Nb, N=3^n$. Then
\be
e^{-V^n(d(\eta))} \sim \frac {1}{(2\pi B)^{3/2}} 
\frac {\eta}{4 \sinh \eta} \exp \left(-\frac 1 {8B}\eta^2\right)
\label{asPotSL2C}
\ee
The Gaussian factor tends to $1$ as $n\mapsto \infty$
because $B\propto 3^n$. 
\subsubsection*{Preliminaries for general $H$}
We note that 
$a_0^0(\Phi)$ and $a_5^0(\Phi)$ 
are positive. By our assumptions,  
$a^0_0 d\Phi $ and $a^0_5(\Phi)d\Phi$ are therefore normalizable measures, and so is their convolution. Let the 
additive constant in $V$ be chosen so that the measure 
$a^0_0\ast a^0_5 d\Phi$ is normalized to one, and let
$ \mathfrak{N}=\int d\Phi e^{-V(\Phi)}$. 
\subsubsection*{$\mathbf{R_+SU(2)}$:}
Elements of $\mathbf{R}_+SU(2)$ have the form $ru$ with $r>0$ and $u\in SU(2)$, but $e^{-V(ru)}$ and $\text{det}(ru)$
are both independent of $u$. Let
\ba
a &=&\int \frac {dr}{r} \ln r \ (a^0_0\ast a^0_5)(ru)\ , \nn \\
a^\prime &=& \int \frac {dr}{r} \ln r\ a^0_0(ru)\ , \nn \\
b &=&\int \frac {dr}{r} \ln^2 r \ (a^0_0\ast a^0_5)(ru)\ , \nn  \\
b^\prime &=& \int \frac {dr}{r} \ln^2 r \ a^0_0(ru)\ ,  \nn \\
A &=& a^\prime + Na \nn \\
B &=& b^\prime + Nb \nn \\
N &=& \frac 1 2 (3^n-1)\ \nn.
\ea
Then the asymptotic behavior of the Higgs potential is as follows
\be
e^{-V^n(ru)} \sim \frac {\mathfrak{N}} {(2\pi B)^{1/2}}
\exp \left( -\frac 1 {2B} \ln^2(re^A)\right) \ .\nn
\ee 
The constant $A$ grows like $N$, but the term proportional $N$ can be transformed away by a wave function renormalization. This is equivalent to modifying the blockspin definition to 
\be
\phi^{n+1}(z+3a_n,z)= Z^{1/2}\phi^n(z+3a_n,z+2a_n)\epsilon
 \phi^n(z+a_n,z+2a_n)^T\epsilon^T \phi^n(z+a_n,z) , \label{PhiZ} \nn
\ee
with a suitable $n$-independent constant $Z$. 
When $A$ stays bounded, the exponential factor tends to $1$ for large $n$, so that the Higgs potential becomes flat.

\subsubsection*{$\mathbf{GL(2,C)}$ and $\mathbf{R_+SL(2,C)}$:}
The asymptotic behavior of $e^{-V(\zeta d(\eta))}$ is independent of the phase of $\zeta$, and combines the factors encountered in the previously considered cases
\be
e^{-V^n(\zeta d(\eta))} \sim \frac{\mathfrak{N}}
{(2\pi B)^{3/2}(2\pi B^\prime)^{1/2}}\ \frac {\eta} {4\sinh \eta} \exp\left( -\frac 1 {8B}\eta^2 - \frac 1 {2B^\prime}\ln^2(|\zeta|e^A)\right). \nn
\ee
with constants $A,B^\prime$ which are of order $N=1$ with $N=\frac 1 2 (3^n-1)$ and $B$ of order $3^n$ and can be computed from the starting values of $V$ similarly as in the other cases.
By a wave function renormalization, $A$ can be transformed to $O(1)$, and the exponential then tends to a constant in the limit. 

\section{Initial Boltzmannian} \label{initialBoltz}
To obtain starting values $a_i^0$ for the RG-flow for general $H$, we need to compute $B^0_F(z,z+1)=e^{S^0_F(z,z+1)}$ for the action
 (\ref{originalAction}).  We retain the definitions (\ref{def_L})
 of $L=L^0,\bar{L}=\bar{L}^0, k=k^0$ and $\bar{k}=\bar{k}^0$ and 
introduce
\ba
\l = \l(z,z+1) &=& \text{det}\big(\phi(z,z+1)\big)\opsi\e\opsi(z)\times\psi\e\psi(z+1)
\ ,  \nn \\
&=& 4  \text{det}\big(\phi(z,z+1)\big) k(z,z+1). \nn \\
\bar{\l} = \bar{\l}(z,z+1) &=& \text{det}\big(\phi(z+1,z)\big)\opsi\e\opsi(z+1)\times\psi\e\psi(z)
\ ,  \nn \\
&=& 4  \text{det}\big(\phi(z+1,z)\big) \bar{k}(z,z+1).\nn
\ea
We will make use of the following identities

\begin{lemma}\label{lemma:7}
For $\Phi\in GL(2,\mathbf{C})$ and 2-component Grassmann variables 
$\psi \neq \opsi$,
$$
\big( \opsi \Phi \psi\big)^2 = -\frac 1 2 [\text{det}\Phi ]\opsi\e\opsi
\times \psi\e\psi
$$
\end{lemma}
{\sc Proof}: The lemma is an immediate consequence of the identities,
\ba \e\phi^T\e^{-1} &=&  [\text{det}\phi]\phi^{-1} \ , \label{eq:ephite}\\ 
\psi_\alpha\psi_\beta = \frac 1 2 \e_{\alpha \beta} \psi\e\psi \ &,& \opsi_\alpha\opsi_\beta = - \frac  1 2 \e_{\alpha \beta} \opsi\e\opsi \ \nn.
\ea

It follows that
\ba
L^2 = -\frac 1 2 \l \  \quad \text{and} \quad 
\bar{L}^2 = -\frac 1 2 \bar{l} \nn \ .
\ea
The expansion of the exponential stops because products of more than two $\psi$'s or more than two $\opsi$'s at any site vanish. As a result one obtains 
\ba 
B_F^0(z,z+1) &=& 
1 + \z L + \bar{\z} \bar{L} -\frac 1 4 \z^2 \l + \z \bar{\z}L\bar L - \frac  1 4 \bar{\z}^2\bar{\l} \nn \\
&& - \frac 1 4 \z^2\bar{\z} \l \bar{L} 
 - \frac 1 4 \z\bar{\z}^2 \bar{\l} L + \frac 1 {16} \z^2\bar{\z}^2 \l\bar{l} \label{notebook_4} \ . 
\ea 
For the coefficients in
 $B^0(z,z+1)=B_F^0(z,z+1)e^{-V(\phi(z,z+1))}$, this yields the following initial values
\ba
\label{eq:z27}
a^0_i &=& b^0_ie^{-V^0} \mbox{ and }\\
b^0_0 &=& 1\nn \\
b^0_1 &=& \z \nn \\
b^0_2 &=& -\z^2 \text{det}\phi \nn \\
b^0_3 &=&  \z \bar{\z}\nn \\
b^0_4 &=& -\z^2\bar{\z} \text{det}\phi \nn \\
b^0_5 &=& \z^2 \bar{\z}^2 \text{det}\phi \text{det}\phi^\ast \ , 
\label{aStart}
\ea
 with the abbreviations $\phi = \phi(z,z+1)$, $V^0=V(\phi(z,z+1))$ and $l=4[\det \phi]  k$ as defined above. 
\section{Integration of the fermions} 
We integrate the Fermi fields $\psi(z)$ and $\bar{\psi}(z)$ attached to sites of the lattice $\Lambda^0$ which are not sites of the block lattice $\Lambda^1$. 
This yields a semi-effective Boltzmann factor $\hat{B}^1(\psi^n, \bar{\psi}^n,\phi)$ which depend on the Fermi fields on the block lattice only, but will still involve the lattice gauge field $\phi$ on all links of the original lattice $\Lambda^0$. 

We admit Higgs fields $\phi(z+1,z)\in GL(2,\mathbf{C})$.
 We will make use of the fact that for $\phi \in GL(2,\mathbf{C})$ eq. 
(\ref{eq:ephite}) is valid
and $-\e=\e^T=\e^{-1}$. For $\phi \in SL(2,\mathbf{C})$, 
$\text{det}\phi = 1$.

\subsection{Basic integration formulae} \label{sec:3.1}
We have two Grassmann variables $\psi_\alpha$, $\alpha=1,2$ attached to each site $z\in \Lambda^0$ and two further Grassmann variables $\opsi_\alpha$. According to the rules for Berezin integrals
\ba \int d^2\psi \ \psi_\alpha = 0\ , \qquad \int d^2\psi \ \label{Berezin}
\psi_\alpha\psi_\beta = \epsilon_{\alpha \beta}\ ,\nn \\
\int d^2\opsi \ \opsi_\alpha = 0\ , \qquad \int d^2\opsi \  \ 
\opsi_\alpha\opsi_\beta = \epsilon_{\alpha \beta}\ . 
\ea

\subsection{The need for odd block lattice spacing} \label{sec:3.2}
One might think of defining an effective Boltzmann factor on a block lattice of lattice spacing $a_1=2$ by integrating out the Fermi fields $\psi(z), \opsi(z)$ at even lattice sites $z$.  However the resulting Boltzmann possesses no logarithm. To see the problem, it suffices to consider the original action (\ref{originalAction}) with Boltzmannian
$B_F^0(z,z+1)e^{-V(\phi(z,z+1))}$ and  with $\z=\bar{\z}=1$. Expanding
the exponentials $B_F(z,z+1)$ into  finite sums of nonvanishing terms and applying eqs.(\ref{Berezin}) one computes $\int d^2\psi(z) d^2\opsi(z)B^0_F(z-1,z)B^0_F(z,z+1).$
One finds that it contains no term of zeroeth order in the Fermi fields. Therefore it possesses no logarithm, and the effective action is not defined. 

\subsection{The semieffective Boltzmannian for block size 3} \label{sec:semieffGeneral}
The problem disappears when we take odd lattice spacing
$a_1=3$. In this case the integration of $\psi(z+1),\opsi(z+1)$ is combined with an integration of the Fermi fields attached to site $z+2$. We start from a general Boltzmannian of the form (\ref{generalBoltzmann}) and compute the semieffective Boltzmannian,
\ba
\hat{B}^1(\psi, \bar{\psi},\phi) &=& \prod_{z\in \Lambda^1} \hat{B}^1_F (z, z+3) \nn \\
\hat{B}^1_F(z, z+3) &=& \int d^2\psi(z+1)d^2\opsi(z+1)d^2\psi(z+2)d^2\opsi(z+2)
\nn \\
&& \qquad
 B^0(z,z+1)B^0(z+1,z+2)B^0(z+2,z+3) \ .  \nn
\ea

We adopt the same choice (\ref{Phi1}) of block spin as before, viz.
\be 
\phi^1(z+3,z)= \phi(z+3,z+2)\epsilon
 \phi(z+1,z+2)^T\epsilon^T \phi(z+1,z) , \label{Phi1shift} \nn
\ee
We retain notations $k$,$L$ introduced before in eqs.(\ref{def_L}),(\ref{def_k}). 

Without loss of generality, we assume that the original action lives on the lattice $\Lambda^0$. This can be achieved by rescaling. 

The semieffective action $\hat{B}_F^1(z,z+3)$ will have the same form
(\ref{generalBoltzmann}), except that the coefficients $a_i^n$ are replaced by coefficients $\hat {a}^1_i(z,z+3)$ which depend on the Higgs field on the fine lattice $\Lambda^0$. More particularly, they depend on 
$\phi_1=\phi(z,z+1), \phi_2=\phi(z+1,z+2)=\phi(z+2,z+1)^\ast, \phi_3=\phi(z+2,z+3)$.
\begin{lemma} \label{lemma:4}
Let us use the abbreviations
 $$\int \dots = \int d^2\psi (z+1)d^2\opsi(z+1) d^2\psi(z+2)d^2\opsi(z+2)\dots $$
and  $$L_1=L^0(z,z+1), \quad L_2=L^0(z+1,z+2), 
\quad L_3=L^0(z+2,z+3) $$
etc., and understand that $L^1=L^1(z,z+3)$ etc. Then
\ba
\int (k\bar{k})_2 &=& 1  \label{1_master}\\
\int L_1(\bar{L} k)_2L_3 &=& L^1 \\
\int \bar{L}_1(L\bar{k})_2\bar{L}_3 &=& \bar{L}^1 \\
\int k_1k_2k_3 &=& k^1 \\
\int \bar{k}_1\bar{k}_2\bar{k}_3 &=& \bar{k}^1 \\
\int (\bar{L}k)_1 L_2 (\bar{L}k)_3 &=& \bar{L}^1k^1 \\
\int (L\bar{k})_1 \bar{L}_2 (L\bar{k})_3 &=& L^1\bar{k}^1 \\
\int (k\bar{k})_1 (k\bar{k})_3 &=& k^1\bar{k}^1 \\
\int (L\bar{L})_1(L\bar{L})_2(L\bar{L})_3 &=&(L^1\bar{L}^1) \label{last_master}
\ea 
and all other integrals of the form $\int X_1Y_2Z_3$ vanish, for $X,Y,Z=L,\bar{L},k,\bar{k}$,$\bar{L}k, ...\quad$.
\end{lemma}    
{\sc Proof:} Using the rules of Berezin integration, one notices that only those factors $X_1Y_2Z_3$ do not integrate to zero for which there are exactly two factors 
$\psi_{\cdot}(z+1)$, two factors $\opsi_{\cdot}(z+1)$, two factors $\psi_{\cdot}(z+2)$ and two factors $\opsi_{\cdot}(z+2)$. Computing their integrals according to the rules of section \ref{sec:3.1} furnishes the above result. {\bf q.e.d.}

Using the lemma \ref{lemma:4}, one can compute 
$$\int \hat{B}^0(z,z+1)\hat{B}^0(z+1,z+2)\hat{B}^0(z+2,z+3)$$
 as a sum of terms $X^1_i=1,L^1,\bar{L}^1,...$ multiplied with coefficients $\hat{a}_i^1(z,z+3)$. We write these coefficients in the form
\be
\hat{a}_i^1(z,z+3)= \sum_{jkl}d^i_{\ jkl}a_j(z,z+1)\bar{a}_k(z+1,z+2)a_l(z+2,z+3). \label{hat_a:1}
\ee  
We use again the notation $a_{\bar{j}}=\bar{a}_j$. With some foresight we chose to define $d^i_{\ jkl}$ so that the second factor $a_{i}(z+1,z+2)$ appears complex conjugated. This is merely a convention, since 
$\sum_{jkl}d^i_{\ jkl}a_j\bar{a}_ka_l=
\sum_{jkl}d^i_{\ j\bar{k}l}a_j{a}_ka_l.$

With this convention one finds that the nonvanishing coefficients $d^i_{\ jkl}$ are equal to $1$, and are as listed at the end of corollary \ref{corollary:1}. 
 We note that
\be
\bar{a}_j(z+1,z+2)= \bar{a}_j[\phi(z+1,z+2)] = a_j[\phi(z+2,z+1)]\ . \label{conj_a} 
\ee

\subsection{The semieffective Boltzmannian for $H=SL(2,\mathbf{C})$}
\label{sec:3.4}
Here we specialize to Higgs fields 
$\phi\in SL(2,\mathbf{C})$. 
In this case, both the initial values (\ref{aStart})
 and the recursion relations (\ref{hat_a:1}) simplify, because $\text{det}\phi = 1$.  

\noindent

\begin{proposition} \label{semieff:SL2C}
Consider the model with holonomy group $H=SL(2,\mathbf{C})$. 
Starting from initial action (\ref{originalAction}), the effective action after the first, and therefore after every RG-step retains the form $($\ref{originalAction}$)$. 
\ba
S^1(\psi^1 ,\bar{\psi^1},\phi^1) &=& \sum_{z\in\Lambda^1}S^1(z,z+3) \nn \\
S^1(z,z+3) &=& \z^1\opsi^1(z)\phi^1(z,z+3)\psi^1(z+3) \nn
 +\bar{\z}^1\opsi^1(z+3)\phi^1(z+3,z)\psi^1(z)\\ && - V^1(\phi^1(z,z+3)) \nn
\ea
In particular 
the semieffective Boltzmannian for $H=SL(2,\mathbf{C})$ after one RG-step has the form 
\ba 
\hat{B}^1(z,z+3)&=& (\z \bar{\z})^2 e^{S^1_F(z,z+3)} 
e^{-V(\phi(z,z+1))-V(\phi(z+1,z+2))-V(\phi(z+2,z+3))} \nn \\
S^1_F(z,z+3) &=& \quad \z^1 \bar{\psi}^1 (z) \phi^1(z,z+3) \psi^1 (z+3) 
+ \bar{\z}^1 \bar{\psi^1} (z+3) \phi^1 (z+3,z) \psi^1 (z)  \nn.
\ea
with $\z^1=-(\z)^2\bar{\z}^{-1}$ and effective Higgs potential $V^1$ as described below.
$a_i^1 =  (\z)^2(\bar{\z})^2 b_i^1 e^{-V^1}$ and $b_i^1$ is 
as in eq. (\ref{aStart}) after having replaced $\z$ by $\z^1$ with $\text{det} \phi =1$.
\end{proposition}
{\sc Proof:} Because $\z$ are constants and 
$\text{det}\phi=1$, the starting coefficients 
$a_i^0$, eq.(\ref{eq:z27}), which determine the original Boltzmannian (\ref{notebook_4}),  become $\phi$ independent 
except for  overall factors $e^{-V^0(\phi(z,z+1))}$ common to all of them which may be pulled out. It follows from eq.(\ref{hat_a:1}) that the coefficients $\hat{a}^1_i$
have the same form as the coefficients in the original Boltzmannian (\ref{notebook_4}) except for the common overall-factor  
$$
(\z \bar{\z} )^2 e^{-V^0(\phi(z,z+1))-V^0(\phi(z+1,z+2))-V^0(\phi(z+2,z+3))}
$$
and replacement of $\z$ by $\z^1$ as defined in the proposition, the factor $(\z \bar{\z} )^2$ remains in the Boltzmannian.
Note that $(\z^1\bar{\z}^1)^2 =(\z \bar{\z})^2$.

Since the original Boltzmannian (\ref{notebook_4}) came from the action (\ref{originalAction}), the same will be true with appropriate adjustments for the semieffective Boltzmannian, as stated in the proposition. 
{\bf q.e.d. }\\
\newline

\section{Consequences of gauge invariance}
\label{sec:gaugeinv}
Here we derive the consequences  of $SU(2)$ gauge invariance,
including eq.(\ref{gaugeInv}). 

\noindent Under $SU(2)$ gauge transformations, $\phi(z,z+1)\mapsto u(z)\phi(z,z+1)u(z+1)^{-1}$. Since $u(z)$ and $u(z+1)$ are both arbitrary elements of $SU(2)$, gauge invariance of $V$ implies that $V(\Phi)=V(u_1\Phi u_2).$ It follows from the parametrization (\ref{dEta})
of $A\in SL(2,\mathbf{C})$ 
 that $V(\zeta A)=V(\zeta d(\eta))$. The remaining statement of eq.(\ref{gaugeInv}) is contained in the following lemma. For later convenience we state it for general functions $V$ on $H$. 

\begin{lemma} \label{lemma:2}\emph{Properties of SU(2)-biinvariant functions}

Suppose $SU(2)\subset H \subseteq GL(2,\mathbf{C})$. 
Let $V$ be a function of $\Phi \in H$, so that $\Phi=\zeta A$ with $0\neq \zeta\in \mathbf{C}$ and $A\in SL(2,\mathbf{C})$ which is $SU(2)$-biinvariant in the sense that 
 $V(\Phi)=V(u_1\Phi u_2)$ for all $u_1,u_2\in SU(2)$. Then 
\be V(\zeta A) =V(\zeta A^{-1}) = V(\zeta \e A^T \e^T). \label{VphiInv} \nn \ee
In particular, for $H=SL(2,\mathbf{C})$ and $d(\eta)=diag(e^{-\eta/2}, e^{\eta/2})$
$$ V(d(\eta))=V(d(-\eta)) \ . $$

\end{lemma}
{\sc proof}: Every element $A \in SL(2, \mathbf{C})$ may be decomposed as $A = u_1d(\eta)u_2 $ with $u_1, u_2\in SU(2)$, $\eta\in \mathbf{R}$, and
$d(\eta) $ as in eq.(\ref{dEta}).

 There exists $w\in SU(2)$ such that $wd(\eta)w^{-1}= d(\eta)^{-1}=d(-\eta)$, viz.
\be w=\left( \begin{array}{ll}
0 & -1 \\
1 & \ 0
\end{array}\right). \label{w}
\ee
It follows that 
\ba V(\zeta A)&=&V(\zeta d(\eta))=V(\zeta wd(\eta)w^{-1})=
V(\zeta d(\eta)^{-1}) = V(\zeta u_2^{-1}d(\eta)^{-1}u_1^{-1}) \nn \\
&=& V(\zeta A^{-1}).\nn \ea
The last equation follows from eq.(\ref{eq:ephite}) with $A$ substituted for $\phi$, since $\text{det} A=1$. {\bf q.e.d.}
%

\section{Integration of the Higgs field for $H=SL(2,\mathbf{C})$}
After the integration of the Fermi fields, we obtained a semieffective Boltzmann factor $\hat{B}^1$  which still depends on the Higgs fields $\phi$ on the links of the original lattice. One sees from the result Proposition \ref{semieff:SL2C} of section \ref{sec:3.4} for 
$H=SL(2,\mathbf{C})$, that the only dependence of $\hat{B}^1 $on $\phi$ other than dependence on the block spin $\phi^1$ is in the factors $e^{-V(\phi(z,z+1))}$. Therefore it only remains to compute the flow of the Higgs potential.
With the choice of blockspin (\ref{Phi}),
the flow of the effective potential is determined by the recursion relation
\be 
e^{-V^{n+1}(\Phi)} = \int d\phi_1\ d\phi_2 \ d\phi_3 \delta(\phi_1\epsilon\phi_2^T\epsilon^T \phi_3\Phi^{-1})
e^{-V^n(\phi_1)-V^n(\phi_2)-V^n(\phi_3)} \ \nn.
\ee
For $\phi\in SL(2,\mathbf{C})$,  
$\epsilon\phi_2^{T}\epsilon^T = \phi_2^{-1}  $ by eq.(\ref{eq:ephite}).

It is easy to see that the biinvariance property of $V$  passes through the recursion relation. Therefore we have lemma \ref{lemma:2} available, with $\zeta=1$, permitting a change of variable so that  
\be 
e^{-V^{n+1}(\Phi)} = \int d\phi_1 d\phi_2 d\phi_3 
\delta(\phi_1\phi_2\phi_3\Phi^{-1})
e^{-V^n(\phi_1)-V^n(\phi_2)-V^n(\phi_3)} \ \nn .
\ee
\begin{lemma}\label{lemma:convol} \emph{(Convolution)}
Under harmonic analysis on $SL(2,\mathbf{C})$, convolution goes into products. 
Thus if the square integrable functions $f_i$ admit expansion 
$$ 
f_i(A)=\int d\chi tr_\chi \left(\tilde{F_i}(\chi)T^\chi(A)\right)
$$
for $i=1...N$  then 
$$
\int dA_1...dA_N f_1(A_1)...f_n(A_n)\delta(A_1...A_NA^{-1}) =
\int d\chi tr_\chi \left(\tilde{F_1}(\chi)... \tilde{F_N}(\chi)T^\chi(A)\right)
$$
Herein, $d\chi $ is the Plancherel measure on $SL(2,\mathbf{C})$, and $T^\chi(A)$ are the representation operators for  representations of the unitary principal series (see Appendix A).  
\end{lemma}
This is well known, and generalizes to the other groups $H$.

Every element $A$ of $SL(2,\mathbf{C})$ may be parametrized as 
\be
A = u_1d(\eta)u_2, \mbox{ with } u_1,u_2\in SU(2), \quad 
 d(\eta)=\left( \begin{array}{ll}
e^{-\eta/2} & 0 \\
0 & e^{\eta/2}
\end{array}\right). \label{dEta}
\ee
It follows from gauge invariance that 
\be V(A)=V(d(\eta))=V(A^{-1}) , \label{gaugeInv} \ee
so that the Higgs potential is a function of a single variable $\eta$. 
We describe the flow of the effective Higgs potential. It is derived by harmonic analysis on the noncompact group $SL(2,\mathbf{C})$, and will involve the principal series representation function 
$ P^\rho(A)={2\sin(\frac 1 2 \eta \rho)}/{\rho \sinh \eta}$
for real $\rho$. 

\begin{theorem} \emph{(Flow of the Higgs potential for $H=SL(2,\mathbf{C})$)} \label{theorem:2}

\noindent
Given the initial Higgs potential $V(\phi)$, define 
\be b(\rho) = \int \sinh^2 \eta d\eta e^{-V(d(\eta))}
\frac {2\sin(\frac 1 2 \eta \rho)}{\rho \sinh \eta} \nn
\ee
for real $\rho$. Then after $n$ RG-steps, 
\be e^{-V^n(d(\eta))} = \frac 1 {8\pi}\int_{-\infty}^{ \infty} \rho^2 d\rho 
[ \z \bar{\z} ]^{2n} 
\left[ b(\rho)\right]^N \frac {2\sin(\frac 1 2 \eta \rho)}{\rho \sinh \eta} \nn
\ee
with $N=3^n$. 

The factor $[ \z \bar{\z}]^{2n}$ show the essential role of the fermions.Because of the recursions relation $\z^1 = (\z)^2 (\bar{\z})^{-1}$ we have $(\z^1)^2 (\bar{\z}^1)^2 = (\z)^2 (\bar{\z})^2$.

\end{theorem}
The result of theorem \ref{theorem:2} is now an immediate consequence of lemma \ref{lemma:convol} and of lemma \ref{lemma:2}.\\

The $SL(2,\mathbf{C})$-model is fairly trivial. The integration of the ``high frequency components'' of the fermions does not affect the flow of the Higgs potential. And the integration of the ``high frequency component'' of the Higgs field $\phi$ does not produce 4-fermion interactions, 6-fermion interactions, or higher.

 This is easy to understand. Because $\psi (z)$ and $\opsi (z)$ are independent variables, they may be transformed independently. The fermionic part of the action will be invariant under the $SL(2,\mathbf{C})$ gauge transformations,
\ba
\psi(z)&\mapsto& A(z)\psi(z) \ , \nn \\
\opsi(z) &\mapsto& \opsi(z)A^\ast(z)\ , \nn \\
\phi(z+1,z) &\mapsto& A(z+1)^{\ast -1}\phi(z+1,z)A(z)^{-1} \ .\nn
\ea
This can be used to transform the parallel transporters in the fermionic action away. Only the Higgs potential is not $SL(2,\mathbf{C})$-invariant.

\section{Effective Boltzmannian  for general $H$}
Here we derive the recursion relations for the 
coefficients $a_i^n$ of the Boltzmann factor for general
 $H$, starting from an arbitrary action of the form
 (\ref{generalBoltzmann}).

Without loss of generality, we assume that the original action lives on the lattice $\Lambda^0$. This can be achieved by rescaling. 

Given the semieffective Boltzmannian of section \ref{sec:semieffGeneral},
 we need to perform the integration over the Higgs fields, keeping the block spin fixed. Switching to variables $\phi_1=\phi(z,z+1), \phi_2=\phi(z+2,z+1)$ (sic!), 
$\phi_3=\phi(z+2,z+3)$ we have to compute the coefficients
\be
a^1_i[\Phi] = \int d\phi_1 d\phi_2 d\phi_3 \hat{a}_i^1(z,z+3)\delta(\phi_1\e\phi_2^T\e^T\phi_3\Phi^{-1}). \nn \label{hat_a:2}
\ee 
Inserting  eq.(\ref{hat_a:1}), and using eq.(\ref{conj_a}) and our notational conventions, eq.(\ref{hat_a:2}) takes the form 
\be
a^1_i[\Phi] = \sum_{jkl} d^i_{\ jkl}\int d\phi_1 d\phi_2 d\phi_3 
a_j[\phi_1]\bar{a}_k[\phi_2]a_l[\phi_3]
\delta(\phi_1\e\phi_2^T\e^T\phi_3\Phi^{-1}). \label{hat_a:3} \nn
\ee 
The coefficients $a_i$ are $SU(2)$-biinvariant functions. Therefore lemma \ref{lemma:2} is applicable to them. We use this to make a change of variable $\phi_2^\prime= \e\phi^T\e^T$ to obtain
\ba
a^1_i[\Phi] &=& \sum_{jkl} d^i_{\ jkl}\int d\phi_1 d\phi^{\prime}_2 d\phi_3 
a_j[\phi_1]a_k[\phi^\prime_2]a_l[\phi_3]
\delta(\phi_1\phi^{\prime}_2\phi_3\Phi^{-1})\nn \\
&=& \sum_{jkl} d^i_{\ jkl} a_j \ast a_k \ast a_l[\Phi] \nn . 
\ea 
Remembering the form of $d^i_{\ jkl}$, this demonstrates Theorem \ref{theorem:3}.
The explicit coefficients are given in theorem \ref{theorem:3}.
Finally we present the solutions of the recursion relations in terms of the initial values $a_j^0$.
\begin{theorem}\emph{(Solution of the flow equations)}
\label{solutionRGeq}

For temporary use we introduce the notation $(A)^{\ast N}=A\ast A\ast ... \ast A$ ($N$ factors) for the $N$-fold convolution product. In this notation
\ba
a_0^n[\Phi]&=& a^0_0*(\bar{a}_5^0\ast a^0_0)^{\ast N}[\Phi], \nn \\
a_1^n[\Phi ]&=& a_1^0 *(\bar{a}_4^0\ast a_1^0)^{\ast N} [\Phi ]
 \ ,\nn \\
a_2^n[\Phi ]&=& a_2^0\ast (\bar{a}_2^0\ast a_2^0)^{\ast N} [\Phi ] \ , \nn \\
 a_3^n[\Phi ]&=& a_3^0 \ast (\bar{a}_3^0 \ast a_3^0)^{\ast N} [\Phi ] \ , \nn \\
 a_4^n[\Phi ]&=& a_4^0 \ast (\bar{a}_1^0\ast a_4^0)^{\ast N} [\Phi ] , \ 
 \nn \\
 a_5^n[\Phi ]&=& a_5^0 \ast (\bar{a}_0^0 \ast a_5^0)^{\ast N} [\Phi ] \  , \nn
\ea 
with $N=\frac 1 2 (3^n-1)$. 

The Fourier coefficients $\tilde{a}^n_j(\Xi)$ obey the same relations, with ordinary products in place of convolution products.
\end{theorem}
The proof is immediate from theorem \ref{theorem:3}.
To use the recursion relations, we need the starting values of the coefficients $a_i(z,z+1)$, and their complex conjugates, that were already given in eqs. (\ref{aStart}).
%
\section{Taking the logarithm} \label{sec:takeLog}
By straightforward computation one deduces
\begin{proposition} \emph{(Effective action from effective Boltzmannian)}\label{proposition:takeLog}

Retaining the assumption $L^2 = -2[\text{det}\phi] k$,
the coefficients $c^n_i$ in the effective action (\ref{generalAction}) are obtained from the coefficients $a^n_j$ in the effective Boltzmannian (\ref{generalBoltzmann})  as follows
\begin{align}
c_0^n=& \text{ln} (a_0^n) \nn \\
c_1^n=& b_1 \nn \\
c_2^n=& b_2 + b_1^2 [\text{det}\phi] \nn \\
c_3^n=& b_3 - b_1 \bar{b}_1 \nn \\
c_4^n=& \bar{b}_4 - \bar{b}_1 b_2 + 2 b_1 b_3 [\text{det}\phi] -2 b_1^2 \bar{b}_1 [\text{det}\phi]\nn \\
c_5^n=& b_5 - b_2 \bar{b}_2 
      + \big( 2 b_1 \bar{b}_4 - \frac{4}{3} b_1^2 \bar{b}_2 \big) 
      [\text{det}\phi] 
      + \big( 2 \bar{b}_1 b_4 - \frac{4}{3} \bar{b}_1^2 b_2 \big)
      [\text{det}\phi^*]\nn \\
      &+ \big( 8b_1 \bar{b}_1 b_3 - 2 b_3^2 -6b_1^2 \bar{b}_1^2 \big) 
       [\text{det}\phi] [\text{det}\phi^*] \nn
\end{align}
where $b_i =\frac{a_i^n}{a_0^n}$ and $\bar{c}_2^n$, $\bar{c}_4^n$ are obtained by complex conjugation.
\end{proposition}
%
\section{Asymptotic behavior after many RG-steps}
We assume that $\z=\bar{\z}=\pm 1$. 

Theorem \ref{solutionRGeq} gives the behavior of the coefficients of the effective Boltzmannian after $n$ steps in the form of a convolution product of $2N+1=3^n$ factors
$a^0_i[\Phi]$, (with $N=\frac{1}{2} (3^n-1)$). 

The central limit theorem gives 
the asymptotic behavior of such convolution products in the limit $N\mapsto \infty$, provided the coefficients 
$a^0_j[\Phi]d\Phi $ are normalizable measures, or negatives of such. Assuming $\z=\bar{\z}=\pm 1$, this is always the case for $a^0_0$ and for $a^5_0$, which enter the formula for the exponentiated Higgs potential 
$a_0^n=e^{-V^n}$ and this is also the case in all cases except for $G=SU(2), H=GL(2,\mathbf{C})$ for the other coefficients.  

We need a central limit theorem for $G$-bicovariant functions on the noncompact groups $H$. The principle from which it flows is as follows. Under appropriate conditions (``good functions $f$'') the Fourier coefficient $\tilde{f}(\Xi)$  for  normalized measures $fd\Phi $ has an isolated absolute maximum at a special representation $\Xi_0$.
The prototype of such a result, which stands behind the central limit theorem on $\mathbf{R}$, is lemma \ref{lemma:5} below. 
 As a result, $\tilde{f}(\Xi)^N$, which is the Fourier coefficient of $f[\Phi]^{\ast N}$, is well approximated by a Gaussian, and therefore 
$f[\Phi]^{\ast N}$ can be accurately computed as inverse transform of a Gaussian. The nontrivial part of the proof of a central limit theorem is the control of errors. To give such a proof is outside the scope of this paper. The proof will be reported in \cite{claudia}. Here we will only present a heuristic derivation, following the above outline. 

In the following subsections, we consider  the exponentiated Higgs potential.
The other coefficients in the effective Boltzmannian can be treated in the same way, {\em provided} the coefficients which enter into the recursion relation are either positive or negative definite. This is true in all cases except for $G=SU(2),H=GL(2,\mathbf{C})$. 

\begin{lemma}\emph{(Fourier transform of normalized measures on $\mathbf{R}$)} \label{lemma:5}
Let $f$ be an integrable and square integrable positive definite function on the real line with $\int f(x)dx=1$. Then the modulus of its Fourier transform $\tilde{f}(k)=\int e^{ikx}f(x)dx $ has a unique absolute maximum at $k=0$ with $\tilde{f}(0)=1$.
\end{lemma}
{\sc Proof}:  
>From the definition of the Fourier transform and from $\int fdx=1$ it follows that
$$ |\tilde{f}(k)|^2 = |\tilde{f}(0)|^2 - \int dx \ dy 
\ [1-\cos k(x-y)]f(x)f(y). $$
The second term is negative semidefinite and for any $k\neq 0$ it can only vanish if $f$ is either a $\delta$-function or a series of $\delta$-functions with supports whose distance is an integer multiple of $2\pi /k$. This is excluded by the assumption that $f$ is square integrable.  Therefore the only absolute maximum is at $k=0$. The assertion $\tilde{f}(0)=1$ is a restatement of $\int f(x)dx=1$. {\bf q.e.d.}

\subsection{The case $H=SL(2,\mathbf{C})$}

We consider the $SU(2)$-biinvariant function
\ba
f(\Phi) = e^{-V(\Phi)} = f(d(\eta)),\nn
\ea
which is positive definite. By our assumptions, the additive constant in $V$ can be so chosen that  $f(\Phi) d\Phi$ is a  normalized measure on $SL(2,\mathbf{C})$.

The Fourier coefficient is
\be
\tilde{f} (\rho)  =  \int d\Phi f(\Phi) P^{\rho}(\Phi^{-1})
    = \int \sinh^2 \eta d\eta f(d(\eta)) \frac{2 \sin \frac{1}{2} \eta \rho}{\rho \sinh \eta} \label{9.1.2} \nn . 
\ee
It is real, obeys $\tilde{f}(\rho)=\tilde{f}(-\rho)$ and has an isolated absolute maximum 1 at 
$\rho=0$. 
This follows from the fact that
$\lim_{\rho\mapsto 0}\frac{2 \sin \frac{1}{2} \eta \rho}{\rho \sinh \eta}=1$, while 
 $|\frac{2 \sin \frac{1}{2} \eta \rho}{\rho\eta}|\leq 1$ and $ |\eta^{-1}\sinh \eta| > 1$ for $\eta \neq 0 $.

If follows from the unique maximum property that 
\ba
\tilde{f} (\rho ) &=& e^{-\frac{b}{2} \rho^2} + ... \nn \\
b&=& -\tilde{f}^{\prime\prime}(0)=\frac{1}{12} \int \sinh^2 \eta d\eta f(d(\eta)) \eta^2 \nn .
\ea
Now set  $B=Nb$, $N=3^n$.
To obtain the asymptotic behavior of the exponentiated effective Higgs potential, we have to back transform
$\tilde{f} (\eta)^N = e^{-\frac{B}{2} \rho^2}$, i.e. compute.
\be e^{-V^n(d(\eta))} \sim \int \frac{\rho^2 d\rho}{8 \pi} \ \frac{2 \sin \frac{1}{2} \eta \rho}{\rho \sinh \eta}  e^{-\frac{B}{2} \rho^2} \nn . 
\ee
The integral evaluates to 
\be
e^{-V^n(d(\eta))} \sim \frac{1}{(2\pi B)^{\frac{3}{2}}}
 \frac{\eta}{4 \sinh \eta } e^{-\frac{1}{8B} \eta^2 }\nn .
\ee
This completes the derivation of the asymptotic behavior of the effective Higgs potential (\ref{asPotSL2C}) 
for $H=SL(2,\mathbf{C})$. 
\subsection{The group $\mathbf{R}_+ SU(2)$}
We specify the representations by $s= i \sigma$ imaginary. $SU(2)$-bicovariant functions $f$ of $ru\in \mathbf{R}_+ SU(2)$ depend only on $r$, and the harmonic analysis amounts to a Mellin transformation (cp. Appendix A), which in turn is related to a standard Fourier transformation,
\be
\label{eq:ftildesu(2)}
\tilde{f} (i\sigma) = \int_0^{\infty} dr r^{i\sigma -1} f(r)\nonumber =\int_{-\infty}^{\infty} dx e^{i \sigma x} f(e^x) \ .
\ee
Let $f$ be positive definite and a good function. Then, by lemma \ref{lemma:5},  $|\tilde f|$ has a maximum at $\sigma=0$ and
\ba
\tilde{f} (i \sigma) = e^{-\frac{b}{2} \sigma^2 + ia\sigma} + \dots \nn
\ea
if $f$ is normalized appropriately.
The coefficients are
\ba 
a&=-i\frac{\partial}{\partial \sigma} f(i\sigma) \big|_{\sigma=0} =
 \int \frac{dr}{r} \ln r f(r)\ , \nn \\
b&=-\frac{\partial^2}{\partial \sigma^2} f(i\sigma) \big|_{\sigma=0} 
 = \int \frac{dr}{r} \ln^2 r f(r). \label{9.2:ab} 
\ea
The exponentiated effective Higgs potential is given by the following convolution product on  $\mathbf{R}_+SU(2),$
\ba
e^{-V^n} = a_0^{*N+1}*a_5^{*N} \nn ,
\ea
with $N=\frac{1}{2} (3^n -1)$, $a_i=a_i^0$, $a_0 = e^{-V}$ and $a_5 = |\det \Phi|^2 e^{-V}$.

We choose the constant in $V$ so, that $a_0 * a_5$ is a normalized measure on $H$ and define $\mathfrak{N}=\int d\phi e^{-V( \phi)}$. 
We have to evaluate
\be e^{-V^n(r)} \sim\int_{-\infty}^{\infty} d\sigma r^{i\sigma} \tilde{a}_0 (i \sigma) \big(  \widetilde{a_0\ast a_5} \big)^N (i \sigma)\ . \nonumber \ee
$a_0*a_5d\Phi$ is convolution of normalizable measures and therefore a normalizable measure itself. Since $a_0$ is positive definite, $a_0*a_5$ is positive definite, implying the  unique maximum property. As a consequence, there are constants $b>0$, $b^\prime >0 $ and $a, a^\prime $ such that
$\tilde{a}_0(i\sigma) = \mathfrak{N}e^{-\frac{b'}{2} \sigma^2 + i\sigma a'}$, $
\widetilde{(a_0*a_5)} (i\sigma) = e^{-\frac{b}{2} \sigma^2 + i \sigma a}$. 
Upon insertion of this and after 
a change of variables to $x=\ln r$, the integral becomes Gaussian and evaluates to 
\be
e^{-V^n(r)} \sim \frac{\mathfrak{N}}{(2\pi B)^{\frac{1}{2}}} \exp \left( -\frac{1}{2B} \ln^2 (r e^A)\right) \nn .
\label{asHiggsRSU2}
\ee
with $B=b^\prime + Nb$, $A=a^\prime+Na$. Their explicit value is determined by the starting Higgs potential $V$ and is obtained by applying eqs.(\ref{9.2:ab}).
\subsection{The group $GL(2,\mathbf C)$} 
This combines aspects of the previous two cases. It is worked out in the same way, using the results of Appendix A. The maximum of $\tilde{f}(\Xi )$, $\Xi=(l,s,\rho)$ is at $l=0, s=0,\rho=0$. The dominance of $l=0$ gives independence of the asymptotic behavior of
 $\exp\left(-V^n(\zeta d(\eta))\right)$ of the phase of $\zeta$.  The group $\mathbf{R}_+SL(2,\mathbf{C})$ is treated in the same way. Here, the quantum number $l$ is absent.
\section{Model with gauge group U(2)} \label{sec:u2}
The results can be extended to the model with gauge group $U(2)$ and $H=GL(2,\mathbf{C})$. To do so, we have to investigate the consequences of an additional local $U(1)$-invariance. Under a $U(1)$-transformation,
\ba \psi(z)&\mapsto& e^{-i\vartheta(z)}\psi(z)\ , \nn \\
\opsi(z+1)&\mapsto &e^{i\vartheta(z+1)}\opsi(z+1)\ ,\nn \\
\phi(z,z+1) &\mapsto& e^{-i\vartheta(z)}\phi(z,z+1)e^{i\vartheta(z+1)} \nn \ . 
\ea  
Local $U(1)$-invariance implies that the Higgs-potential 
$V(\zeta A)$ is independent of the phase of $\zeta$. 
 
The blockspin (\ref{Phi}) is not $U(1)$-covariant, and must be changed by a factor which involves the phase of $\text{det}\phi^n$ to  
\ba
&\phi^{n+1}(z,z+3a_n)=\nonumber\\
&e^{i \text{arg det}\phi^n(z+a_n,z+2a_n)}
\phi^n(z,z+a_n)\epsilon
 \phi^n(z+2a_n,z+a_n)^T\epsilon^T \phi^n(z+2a_n,z+3a_n)
\ . \nonumber  \label{PhiU2}
\ea
It is appropriate to change also the definition (\ref{def_k})  of $k$ to 
\be
k^n(z,z+a_n) = \frac 1 4 e^{i\text{arg det}\phi^n(z,z+a_n)} \opsi\e\opsi(z)\times\psi\e\psi(z+a_n) \nn . 
\ , \label{def_kU2}\\
\ee
When these changes are made, the master integration formulas (\ref{1_master})...(\ref{last_master}) remain valid as they stand. Therefore the recursion relations theorem \ref{theorem:3} for the coefficients $a_j^n$ remain literally valid, and also corollary \ref{corollary:1} remains valid with the understanding that harmonic expansion of $U(2)$-bicovariant functions on $GL(2,\mathbf{C})$ enters.

The initial values (\ref{aStart}) change as a consequence of the changed definition of $k$. In the formulas \ref{aStart}, $\text{det} \phi$ and   $\text{det} \phi^\ast $
have to be replaced by  $|\text{det} \phi |$.

As a consequence, given that $\z=\bar{\z} = \pm 1$, all the  quantities $a^0_j[\phi]d\phi $ are normalizable measures, or negatives of normalizable measure. It follows from this in the same way as for the other models with this property that the model has a  RG-flow from the UV-fixpoint to a IR-fix point. The asymptotic behavior of the Higgs potential is the same as in the $G=SU(2), H= GL(2,\mathbf{C})$  model. 
\section{Outlook}
We are interested in the question how Higgs fields come into an effective field theory at intermediate scales, and what is their geometrical interpretation. 

By definition, effective field theories have an UV-cutoff $M$ with the dimension of a mass, and $M$ can be lowered by a renormalization group transformation \cite{Wilson}.
 Let us assume that the cutoff is introduced in the form of a lattice. 

Our hypothesis is that Higgs fields appear in higher dimensional field theories. At the level of effective theories, they are nonunitary parallel transporters
$\phi$  along links of the lattice in the extra direction or directions. They take their values in a group $H$, typically noncompact, which is larger than the gauge group $G$.   They can arise form unitary parallel transporters, i.e. ordinary gauge theories without Higgs fields, through real space renormalization group transformations.  

Related ideas have been developed in the context of deconstruction \cite{deconstruction}. 

The 1-dimensional model studied in this paper illustrates the mechanism in part. Let us discuss what has to be added and how it involves the conventional 4 dimensions
which have been ignored in the model. We call these the perpendicular directions.

In the 1-dimensional model, the ordinary gauge theory is a fix point. We need a theory in which there are some, albeit arbitrarily small, fluctuations of the length of $\phi$ to start a nontrivial flow.

When the perpendicular directions are taken into account, two things are expected to happen. 
\begin{enumerate}
\item Coarsening in the perpendicular directions, which is involved in the construction of a block spin $\phi$ from unitary parallel transporters $u$ (on the links along the extra direction),
 will produce $\phi$ in the linear span,
 of $G$. Therefore they have fluctuating length of some kind. (Noninvertible $\phi$ will have measure zero.) 
\item At some scale we will be prevented from reducing to a ordinary gauge theory by integrating out the length of $\phi$ (more precisely the selfadjoint factor $p$ in the polar decomposition $\phi=pu$, $u$ unitary),  because otherwise the effective theory will become nonlocal in the perpendicular directions. This will happen when the mass $m$ (of $\pm \rho$) determined by the curvature at the minima of the Higgs potential falls below the cutoff scale $M$. 
\end{enumerate}
In conclusion, there will be a domain of scales where 
the effective theory needs a Higgs field for its locality. In principle, one could also study the mechanisms envisaged here by numerical means.

Let us now turn to the quark sector of the standard model. We give a brief review of results of M. Olschewski and B. Angermann taken from ref. \cite{defect:model}.

 We imagine that all three generations of left handed quarks belong to one irreducible representation space of the holonomy group $H$.  

Ignoring colour, the representation space of $H$ in the standard model is a six-dimensional space $\VL \times \Cdrei$,
 where $\VL$ is the two dimensional representation space 
of the left handed gauge group $G_L=SU(2)\times U(1)$. Similarly, the right handed quarks belong to an isomorphic 6-dimensional representation space $(\VR\oplus \VRtilde) \times \Cdrei$ of $H$,
where $\VR $ and $\VRtilde$ are 1-dimensional representation spaces of the right handed gauge group $G_R=U(1)$
(with $Y=+2/3$ and $Y=-1/3$, respectively)
The Higgs-fields in our sense map between them
\be
\Phi :( \VR\oplus \VRtilde) \times \Cdrei \mapsto \VL \times \Cdrei 
\ee
In a defect model \cite{defect:model}, the Higgs fields $\Phi$ and their adjoints $\Phi^\ast$ are pictured as nonunitary parallel transporters across a defect (brane) which has some extension in a 5-th direction, cp figure 
\ref{bilayered} 
below. 
The left handed quarks live below the defect, and the
 right  handed quarks live above. 

\begin{figure}
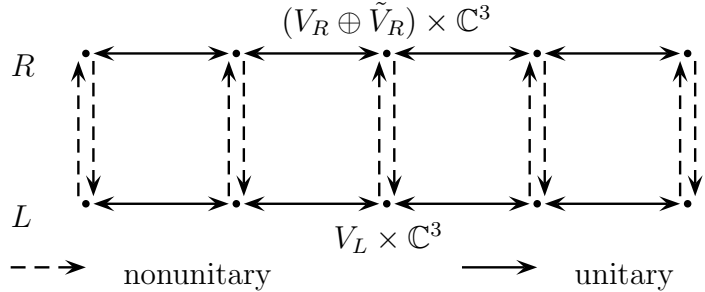


\begin{align*}
\rput[bl](-1.4,0.2){{(V_R \oplus \tilde{V}_R) \times \mathbb{C}^3 }}
\rput[bl](-0.7,-2.65){{V_L \times \mathbb{C}^3  }}
\rput[bl](-5,-0.3){{R }}
\rput[bl](-5,-2.3){{L }}
\text{\pscircle[fillstyle=solid,fillcolor=black](-4,0.0){0.05}
      \pscircle[fillstyle=solid,fillcolor=black](-2,0.0){0.05}
      \pscircle[fillstyle=solid,fillcolor=black](0,0.0){0.05}
      \pscircle[fillstyle=solid,fillcolor=black](2,0.0){0.05}
      \pscircle[fillstyle=solid,fillcolor=black](4,0.0){0.05}
      \pscircle[fillstyle=solid,fillcolor=black](-4,-2){0.05}
      \pscircle[fillstyle=solid,fillcolor=black](-2,-2){0.05}
      \pscircle[fillstyle=solid,fillcolor=black](0,-2){0.05}
      \pscircle[fillstyle=solid,fillcolor=black](2,-2){0.05}
      \pscircle[fillstyle=solid,fillcolor=black](4,-2){0.05}
      \psline[linewidth=.9pt,arrowsize=.20]{<->}(-3.9,0.0)(-2.1,0.0)
      \psline[linewidth=.9pt,arrowsize=.20]{<->}(-1.9,0.0)(-0.1,0.0)
      \psline[linewidth=.9pt,arrowsize=.20]{<->}(0.1,0.0)(1.9,0.0)   
      \psline[linewidth=.9pt,arrowsize=.20]{<->}(2.1,0.0)(3.9,0.0)
      \psline[linewidth=.9pt,arrowsize=.20]{<->}(-3.9,-2)(-2.1,-2)
      \psline[linewidth=.9pt,arrowsize=.20]{<->}(-1.9,-2)(-0.1,-2)
      \psline[linewidth=.9pt,arrowsize=.20]{<->}(0.1,-2)(1.9,-2)   
      \psline[linewidth=.9pt,arrowsize=.20]{<->}(2.1,-2)(3.9,-2)
      \psline[linewidth=.9pt,linestyle=dashed,arrowsize=.20]{->}(-3.9,-0.1)(-3.9,-1.9)  
      \psline[linewidth=.9pt,linestyle=dashed,arrowsize=.20]{->}(-1.9,-0.1)(-1.9,-1.9)  
      \psline[linewidth=.9pt,linestyle=dashed,arrowsize=.20]{->}(0.1,-0.1)(0.1,-1.9)  
      \psline[linewidth=.9pt,linestyle=dashed,arrowsize=.20]{->}(2.1,-0.1)(2.1,-1.9)  
      \psline[linewidth=.9pt,linestyle=dashed,arrowsize=.20]{->}(4.1,-0.1)(4.1,-1.9)  
      \psline[linewidth=.9pt,linestyle=dashed,arrowsize=.20]{<-}(-4.1,-0.1)(-4.1,-1.9)  
      \psline[linewidth=.9pt,linestyle=dashed,arrowsize=.20]{<-}(-2.1,-0.1)(-2.1,-1.9)  
      \psline[linewidth=.9pt,linestyle=dashed,arrowsize=.20]{<-}(-0.1,-0.1)(-0.1,-1.9)  
      \psline[linewidth=.9pt,linestyle=dashed,arrowsize=.20]{<-}(1.9,-0.1)(1.9,-1.9)  
      \psline[linewidth=.9pt,linestyle=dashed,arrowsize=.20]{<-}(3.9,-0.1)(3.9,-1.9) 
}\\
\\
\\
\rput[bl](-3.5,-1.3){{\text{nonunitary} }}
\rput[bl](2.5,-1.3){{\text{unitary} }}
\text{\psline[linewidth=.9pt,linestyle=dashed,arrowsize=.20]{->}(-5,-1)(-4,-1)
      \psline[linewidth=.9pt,arrowsize=.20]{->}(1,-1)(2,-1)}
\\
\\
\end{align*}
\caption{ Bilayered membrane}

 \label{bilayered}

\end{figure}

There will be a Higgs potential $V(\Phi)$.
We imagine that the 
Higgs potential is to be determined as the result of a renormalization group flow, similarly as in the very simplified model of this paper. Unfortunately we are unable to say, for now, where the flow starts. In the simplified model, the asymptotic behavior of the Higgs model is quite insensitive to where the flow starts. 
This reflects universality.   

If the Higgs potential could be computed,
 the quark masses could be determined, as explained in section \ref{sec:masssplit}. The value of $\Phi$ at the minimum of the Higgs potential will contain the information that goes into the quark masses.

Writing $\varphi$ for the conventional Higgs doublet,
  \begin{equation}
   \label{vphi}
   \varphi      = \left(
\begin{array}{l}
 \varphi_0 \\ \varphi_+
\end{array}
 \right)  \qquad  
   \tilde{\varphi} = \left(
\begin{array}{l}
 -\bar{\varphi}_+ \\ \bar{\varphi}_0
\end{array}
 \right)  \qquad
   \left.   \begin{array}{l} \varphi:\VR \to \VL \\ \tilde{\varphi} :\VRtilde \to \VL   \end{array}   \right.    
  \end{equation}
the Higgs parallel transporter $\Phi$ through the defect will have the form
  \begin{equation}
   \label{HiggsParallelRL}
   \Phi = (\varphi \MU,\tilde{\varphi} \MD) \; : \; (\VR \oplus \VRtilde) \times \Cdrei \to \VL \times \Cdrei  \ .
  \end{equation}
Neglecting fluctuations of the Higgs field $\Phi$ around the minimum of the Higgs potential, $\MU$ and $\MD$ will be constant $3\times 3$ matrices
\be
\MU, \MD: \Cdrei \mapsto \Cdrei . 
\ee
The matrices $\MU , \MD$ can be diagonalized through a biunitary transformation,
\ba \MU &=& A_L^\ast m_UA_R, \quad m_U=diag(m_u,m_c,m_t) \\
\MD &=& B_L^\ast m_D B_R, \quad m_D=diag(m_d,m_s,m_b).
\ea
Herein, $m_u,...$ are the quark masses,
and $A_L,B_L, A_R, B_R$ are unitary. By a basis change, $A_R,\ B_R$ can be transformed away. The Kobayashi Maskawa matrix is $C=A_LB^\ast_L$.

Consider now the elements of the holonomy group $H_{tot}$
of the whole theory, including the boundaries of the defect which were not modeled in our 1-dimensional theory.

The elements of $H_{tot}$ 
are the parallel transporters along paths which may pass through the defect, possibly several times forth and back, and their inverses. The parallel transporters along pieces of path above the defect will be of the form $U_R{\mathbf 1}$, where $\mathbf{1}$ is the $3\times 3$ unit matrix, and $U_R$ are diagonal $2\times 2$  matrices
whose nonvanishing entries are representation operators of $U(1)$, and the parallel transporters along pieces of path below the defect will be unitary matrices $U_L\mathbf{1}$ with    
\be U_L = \left( \begin{array}{ll} 
U_{11} & U_{12}\\
U_{21} & U_{22} \end{array} \right) \in SU(2)\times U(1) \ .
\ee
Going to unitary gauge $\varphi_0=\rho, \varphi_+=0$,
the parallel transporter along a closed path which passes once across the defect, forth and back, will be of the form
\be U_R\Phi^\ast U_L\Phi = U_R 
\left(\begin{array}{ll}
m_U & 0 \\
0 & m_D
\end{array}\right)
\left(\begin{array}{ll}
U_{11}\mathbf{1} & CU_{12} \\ C^\ast U_{21}& U_{22}\mathbf{1}
\end{array}\right)
\left(\begin{array}{ll}
m_U & 0 \\
0 & m_D
\end{array}\right) \rho^2
\ee
where $C$ is again the Kobayashi-Maskawa matrix.
The difference of the quark masses is responsible for the fact that the holonomy group $H$ is larger than the unitary gauge group. It involves nonunitary matrices and is therefore noncompact.
\section*{Acknowledgment}
It is a pleasure to thank B. Angermann, M. Olschewsky, F. Neugebohrn, T. Pr\"ustel, M. de Riese and M. R\"ohrs for discussions. C.L wishes to thank the Deutsche Forschungsgemeinschaft for financial support through the Gradu\-ierten\-kolleg {\em Zuk\"unftige  Entwicklungen in der Teilchenphysik}.

\section*{Appendix A: Harmonic analysis}
Here we detail the expansion of $SU(2)$-biinvariant
functions $f(\Phi)$ on $H$ in spherical functions $P^\Xi(\Phi)$. They are associated with unitary representations $\Xi$ of $H$. The general expansion formula reads
\be
f[\Phi] = \int d\Xi \tilde{f}(\Xi )P^\Xi(\Phi ) \nn ,
\ee
where d$\Xi$ is the restriction of the Plancherel measure on $H$  to representations which contribute to the expansion of $SU(2)$-biinvariant functions. 

Denoting by $\overline{\Xi}$ the complex conjugate representation, the complex conjugate function $\bar{f}$ expands according to
\be \bar{f}[\Phi] = \int d\Xi \tilde{f}(\Xi )P^{\overline{\Xi}}(\Phi ) \nn . \label{FT}
\ee

\subsection*{Harmonic analysis on $SL(2,\mathbf{C})$}
We use the parametrization (\ref{dEta}) of elements $SL(2,\mathbf{C})$
in terms of variables $u_1,u_2\in SU(2)$ and $\eta \in \mathbf{R}$. 

Integration with the Haar measure in these variables can be written as
\be
\int d\phi f(\phi )  = \int du_1\ du_2 \sinh^2 \eta d\eta  f(u_1d(\eta)u_2) \label{Haar}
\ee
where $du$ is normalized Haar measure on the group $SU(2)$. The formula involves a redundant integration over the commutant of $d(\eta)$ in $SU(2)$.

We wish to use harmonic analysis on the group $SL(2,\mathbf{C})$ to determine the RG-flow of functions $f$ like the exponentiated Higgs potential $V$. In case $H=SL(2,\mathbf{C})$. They are
functions on $H=SL(2,\mathbf{C})$. We assume that $V(d(\eta))$ increases sufficiently fast at large $|\eta|$ to ensure that these functions are square integrable on the group. This ensures \cite{GelfandShilov, Wallach} that it can be expanded in representation functions of the principal series of unitary representations of $SL(2,\mathbf{C})$. 

The representations of the Lorentz group $SO(3,1)$ in the principal series are labeled by $\chi=(l,\rho)$, where $l$ labels  a representation of $SO(2)$, $l=0,\pm1,\pm2,...$  and $\rho$ is a real number. $SL(2,\mathbf{C})$ is the twofold covering of $SO(3,1)$. Its principal series representations are labeled in the same way, except that $l=0,\pm\frac 1 2 , \pm 1,...$. The representation operators $T^\chi(A)$ act on Hilbert spaces $\H^\chi $. The expansion of a square integrable function on the group reads 
\footnote{$l=\frac 1 2 m$ in R\"uhls notation \cite{Ruehl}}
\be 
f(A)=\int d\chi tr_\chi \left(\tilde{F}(\chi)T^\chi(A)\right)\nn
\ee
where $d\chi$ is the Plancherel measure on $SL(2,\mathbf{C})$
\be
\int d\chi \dots = \sum_l\ \int_{-\infty}^{\infty}(l^2 + \rho^2) \frac {d\rho}
{8\pi} \dots \nn ,
\ee   
$tr_\chi$ is the trace over $\H^\chi$, and $\tilde{F}(\chi)$ is an operator in $\H^\chi$. 
The inversion formula reads
\be
\tilde{F}(\chi) = \int_{SL(2,\mathbf{C})} d\phi f(\phi)T^\chi (\phi^{-1}) \nn. 
\ee
We apply this to the functions $f(A)$ which are biinvariant in the sense that $f(u_1Au_2)=f(A)$ for all $u_1,u_2\in SU(2)$. A great simplification results. The representation space $\H^\chi$ decomposes into irreducible representations of $SU(2)$ as follows \cite{Ruehl, Dobrev}. Every irreducible representation of $SU(2)$ occurs at most once, and the $2k+1$-dimensional representation occurs if $k\geq |l|$, $k-l$ integer. In particular, the trivial representation of $SU(2)$ occurs only for $l=0$. Using this decomposition, one introduces in $\H^l$ the socalled canonical basis $\{ |k,m> \}$ where $k \geq |l|$ fixes the $SU(2)$-representation, and $m=-k,...,k$ in integer steps. $m=0$ if $k=0$.
The representation functions in the canonical basis are known, in particular
\cite{Ruehl}
\be
P^\rho(A) = <0,0|T^{(0,\rho)}(A)|0,0> = \frac {2\sin (\frac 1 2 \eta \rho)} {\rho \sinh \eta}
\label{d00}
\ee
for $A=u_1 d(\eta) u_2$, $u_i\in SU(2)$. 
\begin{lemma} \label{lemma:3}
If $f(A)=f(u_1Au_2)$ for all  $u_1,u_2\in SU(2)$ then 
\be
<k,m|\tilde{F}(\chi)|k^\prime,m^\prime>  = \delta_{l0}\delta_{k0}\delta_{k^\prime 0}\tilde{f}(\rho) \nn
\ee
for some $\tilde{f}(\rho) \in \mathbf{C}$. 
\end{lemma} 
This is well known, cf. \cite{Ruehl}.

Using the lemma, the expansion simplifies, 
\be 
f(A) = \int_{-\infty}^{\infty} \frac {\rho^2 d\rho}{8\pi} \tilde{f}(\rho)
<0,0|T^{(0,\rho)}(A)|0,0> \nn
\ee
The normalization of the Plancherel measure depends on the normalization of the Haar measure. We use the Haar measure (\ref{Haar}) in the parametrization $A=u_1d(\eta)u_2$.

In summary, the relevant representations $\Xi=(\rho)$ are labelled by one real parameter  $\rho$, 
\be \int d\Xi \dots  = \int_{-\infty}^{\infty}  \frac {\rho^2 d\rho}{8\pi} \dots\  \nn,
\ee
and the spherical function $P^\rho$ is given by eq.(\ref{d00}). Taking its complex conjugate, we see that 
the complex conjugate representation is $\overline{\Xi}=(-\rho)$. 

 Since the $u_1,u_2$ integrations are trivial, the inversion formula simplifies to  the form used in Corollary \ref{corollary:1}.

Let us note that the inversion formula can also be obtained by elementary means, using the orthogonality of the $\sin $-function, 
\be 
\int_{-\infty}^{\infty} d\rho \ \sin \frac 1 2 \eta \rho 
\ \sin \frac 1 2 \eta^\prime \rho
 = 2\pi [\delta(\eta - \eta^\prime) - \delta (\eta + \eta^\prime)] \nn ,
\ee
and $\tilde{f}(-\rho)=\tilde{f}(\rho)$. In this way normalization factors can be fixed. The group theoretical interpretation is needed to make lemma \ref{lemma:convol}
available, though, in order to convert convolutions into products. . 

In applications we use 
the $\delta$-function on $SL(2,\mathbf{C})$ with defining property
$$
\int d\phi \delta(\phi)f(\phi ) = f(\mathbf{1}).   
$$
Note that its normalization depends on the normalization of the Haar measure.
\subsection*{Harmonic analysis on $H=\mathbf{R}_+ SU(2)$}
Elements $\Phi \in H=\mathbf{R}_+ SU(2)$ have the form
$\Phi = r u$, with positive real $r$ and $u\in SU(2)$. $SU(2)$-biinvariant functions $f(\Phi)$ depend only on $r$. Therefore harmonic analysis 
simplifies to a Mellin transform,
\be
f(ru) = \int_{-i\infty}^{i\infty} 
\frac {ds}{2\pi i} r^{s} \tilde{f}(s) ,\qquad
 \tilde{f}(s) = \int_0^\infty dr r^{-s-1} f(ru) \nn .
\ee
Thus, the representations $\Xi$ of $H$ which appear in the expansion of 
$SU(2)$-biinvariant functions, are specified by pure imaginary numbers 
$s$, and the Haar measure and spherical function are
\be \int d\Xi \dots = \int  \frac {ds}{2\pi i} \dots\ , \qquad
P^\Xi(ru)=r^{-s} \nn .
\ee
The complex conjugate representation is $\overline{\Xi}=(s)$. 
\subsection*{Harmonic analysis on $H=GL(2,\mathbf{C})$}
Elements of of $GL(2,\mathbf{C})$ have the form $\Phi=\zeta A$, 
$0\neq \zeta \in \mathbf{C}$, $A\in SL(2,\mathbf{C})$. Note that the pair $(-\zeta, -A)$ determines the same element as $(\zeta, A)$, and that $SU(2)$-biivariant functions $f$ obey $f(-\zeta A)=f(\zeta A)$ because
 $(\mathbf{-1})\in SU(2)$. 

It follows that we have to combine harmonic analysis on 
$SL(2,\mathbf{C})$ with harmonic analysis on the multiplicative group
$\mathbf{C}\setminus \{ 0 \}$. 

Consider functions $g$ of $\zeta \in \mathbf{C}\setminus \{ 0 \}$ with $g(-\zeta)=g(\zeta)$. Let $\zeta = e^{i\vartheta /2}r,\ r>0$. Then
\ba
g(\zeta) &=& \sum_{l\in\mathbf{Z}}\int \frac {ds}{2\pi i}
 e^{il\vartheta} r^{s} \tilde{g}(l,s) \nn ,   \\
\tilde{g}(l,s) &=& \int_0^{2\pi} \frac {d\vartheta}{2\pi} dr
 e^{-il\vartheta} r^{-s-1} g(e^{i\vartheta/2}r) \nn .
\ea
Combining with harmonic analysis on $SL(2,\mathbf{C})$, we see that the representations $\Xi$ of $GL(2,\mathbf{C})$ which appear in the expansion of $SU(2)$-biinvariant square integrable functions are labelled by 
$\Xi=(l,s,\rho)$, with $l\in \mathbf{Z}$, $s$ imaginary, and 
$\rho $ real, with Plancherel measure and spherical function
\ba
\int d\Xi \dots &=& \sum_{l\in \mathbf{Z}}\int_{-i\infty}^{\infty} 
\frac {ds}{2\pi i} \int_{-\infty}^\infty
 \frac {\rho^2d\rho}{8\pi}\dots \nn ,\\
P^\Xi(e^{i\vartheta /2}rA) &=& e^{il\vartheta}r^s P^\rho(A)\nn ,
\ea
where $P^\rho(A)$ is the spherical function on $SL(2,\mathbf{C})$. 

The complex conjugate representation is labelled by 
$\overline{\Xi}=(-l,-s,-\rho)$. 

\end{document}